# Fluid approach of current-driven Langmuir waves associated with type III radiation and whistler waves: Relevance to PSP and other solar wind observations

Konrad Sauer[1] and Kaijun Liu[2]

[1] *Max-Planck Institute of Solar System Research, Göttingen, Germany*
[2] *Southern University of Science and Technology, Shenzhen, Guangdong, China*

**Abstract** A theoretical model on the basis of fluid-Maxwell equations for an electron-ion plasma is presented which describes the conversion of current-driven Langmuir waves into type III radiation whereby simultaneously an excitation of whistler waves may occur. In contrast to the classical approach of Ginzburg and Zhelezniakov (1958) after which beam-excited Langmuir waves in a two-step process are converted in electromagnetic radiation, the presented mechanism works without any instability and wave coalescence. Rather the electric field oscillations at the electron plasma frequency can be triggered by different realisations of the driving current, e.g. by the (uncompensated) net current of the strahl at t=0 in a core-strahl plasma or by given current variations which may represent different situations in space, as shocks, magnetic switch-backs etc.. A linearized system of equations is used to describe the mode coupling occurring at oblique propagation between the mostly electrostatic Langmuir wave and the adjacent electromagnetic left-hand polarized (L) wave. The simplicity of the fluid model allows without great effort the parameters of the current profiles to be varied and thus to simulate a wide range of possible experimental conditions. Measurements of Langmuir waves, type III radiation and whistler waves on board various satellites in the solar wind, and in particular some of the recent results of the Parker Solar Probe are interpreted in the light of the theoretical model presented. For the case of the uncompensated strahl, the fluid approach is confirmed by fully kinetic PIC simulations. One comparison is shown in the Appendix.

## 1. Introduction

The presented theoretical approach of explaining the origin of the solar radio emission (type III radiation) and its association with Langmuir and whistler waves has a relatively long history. It goes back to first studies by Sauer and Sydora (2012) in which the effects of mode coupling have been considered by means of PIC simulations. In using the 1D electromagnetic code, the simulation of the classical beam-instability (in case of quasi-parallel propagation) has shown that besides growing Langmuir waves in the range of resonant wave numbers k given by $\omega_e = kV_b$ ($\omega_e$: electron plasma frequency, $V_b$: beam velocity), an excitation of the left-handed electromagnetic (L) wave takes place where the Langmuir wave and the L wave cross each other. Detailed analysis brought the clear conclusion that this effect cannot be reconciled with the two-step model of Ginzburg and Zheleszniakov (1958). Simultaneously, whistler waves were excited at frequencies below $\Omega_e/2$ ($\Omega_e$: electron cyclotron frequency), which could not be directly connected to the beam due to a lack of resonance. A new element came into the considerations with the concept of current-driven Langmuir waves (Sauer and Sydora, 2015), which are not caused by an instability, but by an initial current. This made it easy to show that a stable plateau distribution allows Langmuir waves to exist over a long period of time after the instability has dissipated. Later, a fluid approach, very similar to that presented in this paper, has been developed (Sauer et al., 2019) which describes the coupling between the current-driven Langmuir wave and the adjacent L wave in an over-dense plasma ($\Omega_e/\omega_e < 1$).

The diversity of observations by the Parker Solar Probe (PSP), in particular the diverse spectra of whistler waves, stimulated us to extend our focus on a broad spectrum of electromagnetic waves which is basically contained in the system of cold fluid and Maxwell equations. In the present model, a plasma is considered that consists of a core and a strahl distribution which provides the initial current. By linearizing the basic equations, it is possible to calculate the response of the current to the complex two-electron plasma whereby the ions are considered to be fixed. Besides the strahl, the origin of currents triggering electromagnetic waves can be very different. They may occur in the vicinity of density and the magnetic field discontinuities, such as shocks and boundaries of magnetic clouds. Moreover, magnetic switch-backs and holes (Agapitov et al., 2020) also come into consideration.

The simultaneous generation of high-frequency waves near $\omega_e$ and whistler waves with $\omega \leq 0.1 \Omega_e$ is an essential feature in our theory of current-driven waves. Observations of this kind can many times be found in the literature, e.g., in Moullard et al. (1998, 2001) and have recently aroused new interest through measurements on PSP concerning their origin, see e.g., Jagarlamudi et al. (2021). However, the variety of recent measurements of whistler waves on PSP alone poses a challenge for the predictions of our theoretical model. It will be shown that a number of measured whistler wave spectra can be explained in particular by the variation of the propagation angle.

The outline of the paper is the following: After presenting in Section 2 the fluid-Maxwell equations for a core-beam plasma and the dispersion of the associated waves, subsequently the excitation of Langmuir waves and their transformation into electromagnetic waves is investigated for the case where the strahl current at time t=0 is not compensated by a corresponding displacement of the core. This switch-on configuration represents the excitation over a wide frequency range and has the advantage that a comparative PIC simulation can be easily realized without major modifications to the code. In Section 4, the situation is adapted more closely to real space conditions by specifying temporal variations for the current that are adapted to the processes during beam instability or are related to possible current changes during magnetic holes or switch-backs etc. The Summary and Discussion follows in Section 5. In the Appendix, the power spectra of the electric and magnetic fields

## 2. Fluid approach and related waves

In this section, the basic mechanism of excitation of Langmuir waves and their conversion into electromagnetic radiation is described. Essential element is the excitation of Langmuir waves by temporal variations of the electron current. In case of spatially homogeneous conditions (wave number k=0) an initial current drives Langmuir oscillations at the electron plasma frequency, as shown by Sauer and Sydora (2015). Although the simplest conceivable situation is the drift of electrons with respect to ions, with a view to later applications to space observations, in Section 3. an electron plasma of core electrons with a small fraction of drifting electrons, hereafter called the strahl, is considered. This is followed in Section 4. by model calculations with (prescribed) current profiles which are intended to mimic the situation of beam instability.

A simple one-dimensional model is used in which the magnetic field **B₀** lies in the x-z plane and is inclined by the angle θ relative to the x-axis which is taken as the direction of the wave propagation. Ions are treated as an immobile background. Two electron components corresponding to the core and strahl distributions are assumed in order to match real conditions in the solar wind. Both components are considered cold. The basic equations consist of the continuity equation and the equation of motion for both electron populations, combined with the Maxwell equations. The index c and s stand for the core and the strahl, respectively.

For our purposes it is sufficient to use the linearized equations which are written in normalized form as follows

*Core population*

$$\frac{\partial V_{cx}}{\partial t} = -G[E_x + V_{cy}B_{z0}] \qquad (1)$$

$$\frac{\partial V_{cy}}{\partial t} = -G[E_y + (V_{cz}B_{x0} - V_{cx}B_{z0})] \qquad (2)$$

$$\frac{\partial V_{cz}}{\partial t} = -G[E_z - V_{cy}B_{x0}] \qquad (3),$$

*Strahl population*

$$\frac{\partial V_{sx}}{\partial t} = -G[E_x + V_{sy}B_{z0}] \qquad (4)$$

$$\frac{\partial V_{sy}}{\partial t} = -G[E_y + (V_{sz}B_{x0} - V_{sx}B_{z0})] \qquad (5)$$

$$\frac{\partial V_{sz}}{\partial t} = -G[E_z - V_{sy}B_{x0}] \qquad (6),$$

*Maxwell equations*

$$\frac{\partial E_x}{\partial t} = (n_{c0}V_{cx} + n_{s0}V_{sx})/G \qquad (7),$$

$$\frac{\partial E_y}{\partial t} = (n_{c0}V_{cy} + n_{s0}V_{sy})/G - \varkappa B_z \qquad (8),$$

$$\frac{\partial E_z}{\partial t} = (n_{c0}V_{cz} + n_{s0}V_{sz})/G + \varkappa B_y \qquad (9),$$

$$\frac{\partial B_y}{\partial t} = +\varkappa E_z \qquad (10),$$

$$\frac{\partial B_z}{\partial t} = -\varkappa E_y \qquad (11).$$

Here, $\varkappa = kc/\omega_e$ is the normalized wave number, t is in units of $\omega_e^{-1}$. and $n_{s0}$ are the equilibrium electron densities of the core and the strahl normalized to the total electron density $n_0 = n_{c0} + n_{s0}$, respectively. $G = \Omega_e/\omega_e$ is the ratio between $\Omega_e$ and $\omega_e$. The electron velocity components are normalized to c, the light velocity. The electric field is in units of $E_0 = B_0 c$ and the magnetic field perturbations are normalized to the undisturbed magnetic field $B_0$. Since the density perturbations are absent in the equations for the electric field, the continuity equation has not been listed explicitly.

As a first step, the dispersion relation for a simple electron-ion plasma ($n_{s0} = 0$) is derived. Using the periodic function $\sim\exp\{i\omega t\}$ for all quantities in the fluid-Maxwell equations above, one obtains (using $y=\omega/\omega_e$) the relation $D(y,\varkappa)=0$, where the dielectric function $D(y,\varkappa) = C_0+C_1 z+C_2 z^2+C_3 z^3+C_4 z^4$ is a polynomial of fourth order in $z=y^2$ with the coefficients $C_0$, $C_1$, $C_2$, $C_3$ and $C_4$ which depend on the wave number x, the propagation angle $\theta$, and the ratio G, see also Sauer and Sydora (2012) and Sauer et al. (2019).

In Figure 1 examples are shown for under-dense plasmas with a) G=0.025 and b) G=0.1 using a propagation angle of $\theta = 15^0$. In the top two panels only the two high-frequency electromagnetic modes, the right- (R) and left-hand polarized (L) mode and the Langmuir mode around the plasma frequency are shown. The whistler mode, which is nearly the same for both values of G, is shown in panel c). A remarkable dispersion property in case of oblique wave propagation is the mode coupling between the Langmuir wave and the adjacent electromagnetic L mode, which is marked by the red dashed circles in the panels a) and b). This signature is a key element in our model of electromagnetic wave generation.

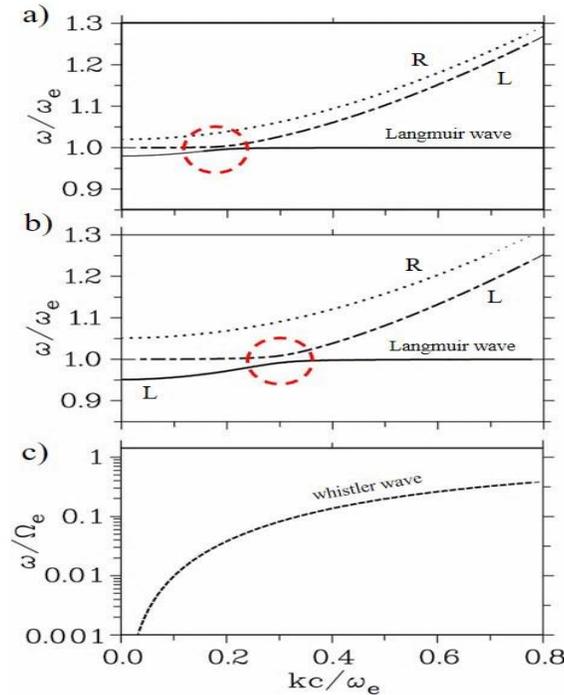

**Figure 1.** Dispersion of the Langmuir wave and adjacent R- and L- electromagnetic modes for $\theta=15^0$ and two values of $G=\Omega_e/\omega_e$: a) G=0.025, b) G=0.1. In case of oblique propagation, mode splitting takes place due to the coupling of the Langmuir (Langmuir) and the left-handed eletromagnetic wave (L) which cross each other; see the encircled region. With the increase of G, the crossing point shifts to larger wave numbers. Panel c) represents the whistler dispersion which for G<<1 doesn't depend on G.

Since the cut-off frequencies for G < 1 are given by $\omega/\omega_e = 1 \pm G/2$, the mode crossing point shifts to smaller wave numbers if G decreases. For example, for G=0.1 the corresponding wave number is $kc/\omega_e = 0.3$ and it moves to $kc/\omega_e \sim 0.15$ if G decreases to G=0.025.

## 3. Electromagnetic wave generation by a non-compensated strahl current

In this section, the fundamental fluid equations for a core-beam plasma are solved for the case where, at the beginning of the temporal evolution (i.e., at t=0), the beam current is not compensated by a corresponding displacement of the core electrons. Although this initial situation does not initially appear to be directly related to the real conditions for the generation of electromagnetic radiation by electron beams, this configuration is important for the context of our model. For example, in the work of Yao et al. (2022), the plasma emission in a 2.5D beam plasma system was investigated under the assumption that, at time t=0, an electron distribution exists that corresponds to an uncompensated current resulting from magnetic reconnection. However, the interpretation of the results is made without reference to the mechanism of current-driven Langmuir waves and their possible transformation into radiation. Evidence of the role of net currents resulting from beam activation can be found in the work of Thurgood and Tsiklauri (2015). While the work points to current-driven electric fields with vanishing wavenumber (k=0), the processes associated with beam instability subsequently become the focus of further investigations, which is certainly also due to the insufficient resolution in the wavenumber range.

For us, an important aspect of analyzing the effect of an uncompensated current concerns the easy possibility of comparing the results of the fluid approach with those of a PIC simulation using the same conditions. The abrupt switch-on process offers the advantage of providing a broad spectrum of excitation frequencies and thus facilitating an overview of possible 'resonances', which as we will see is manifested by the simultaneous occurrence of Langmuir and whistler waves.

### 3.1 Langmuir oscillations at parallel propagation
First, we consider pure Langmuir oscillations in the case of parallel propagation. Extending the equations (4) and (5) to a core-strahl plasma, then applies

$$\frac{dV_{cx}}{dt} = -GE_x \qquad (12)$$

$$\frac{dV_{sx}}{dt} = -GE_x \qquad (13)$$

$$\frac{dE_x}{dt} = +\frac{1}{G}(n_{c0}V_{cx} + n_{s0}V_{sx}) \qquad (14)$$

This results in the simple equation of oscillations

$$\frac{d^2}{dt^2}E_x + E_x = 0 \qquad (15)$$

with the solution

$$E(t) = E_{s0} \cdot \sin t, \quad (16)$$

where the amplitude $E_{s0}$ is given here by the strahl current according to

$$E_{so} = n_{s0} V_{s0}/G. \quad (17)$$

The corresponding velocities of the core and strahl electrons are determined by the relations

$$V_{cx}(t) = E_{s0} (\cos t - 1) G \quad (18)$$

and

$$V_{sx}(t) = V_{s0} + V_{cx}(t) \quad (19)$$

For the following example, a strahl with the density $n_{so} = 0.05$ is considered. With regard to the strahl velocity, for an adaptation to the results of PIC simulations presented in the companion paper (Sauer and Liu, 2025), a (fictitious) thermal velocity of the core ($V_e$) is introduced. Furthermore, a ratio $V_e/c = 0.05$ is taken. This allows the strahl velocity to be expressed in units of $V_e$ instead of c. With an assumed strahl velocity of four times the thermal velocity ($V_s = 4V_e$) and $G = 0.1$, equation (17) gives an amplitude of the Langmuir oscillations of $E_{s0} = 0.1$. For these parameters, the temporal evolution of the core velocity $V_{cx}$, the strahl velocity $V_{sx}$ and the electric field is illustrated in Figure 2. As seen on the left, the electric field $E_x(t)$ (in units of $E_0 = cB_0$) oscillates symmetrically between -0.1 and +0.1. The particles' velocities, on the other hand, only oscillate from the starting point in the direction of smaller values: $-0.4 \leq V_{cx}/V_e \leq 0$ for the core electrons, $3.6 \leq V_{sx}/V_e \leq 4$ for the strahl electrons.

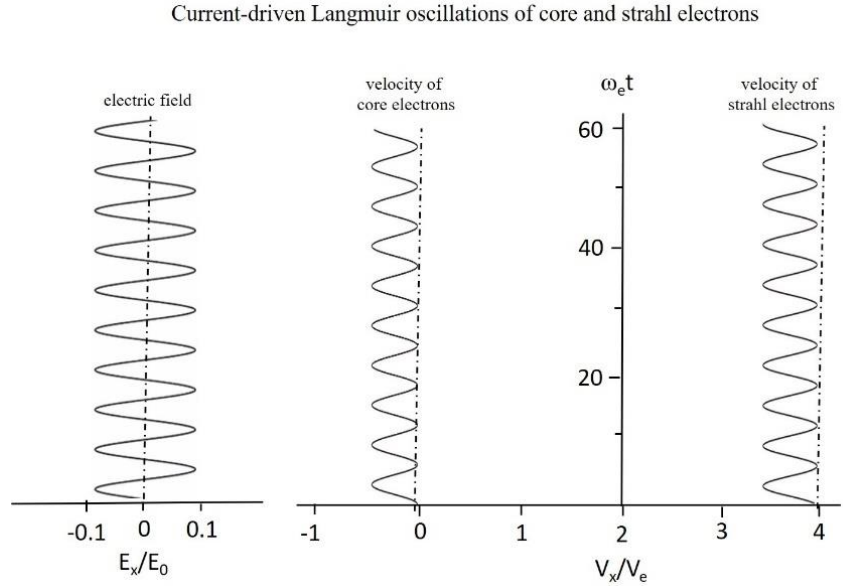

**Figure 2.** Current-driven Langmuir oscillations in case of parallel propagation ($\theta = 0^0$). Left: electric field $E_x(t)/E_0$, middle: velocity of core electrons $V_{cx}(t)/V_e$, right: velocity of strahl electrons $V_{sx}(t)/V_e$. The driving current is caused by the strahl with the density $n_{s0} = 0.05$ and velocity $V_s = 4V_e$ where a core thermal velocity of $V_e = 0.05/c$ was taken. The further parameter is $G = \Omega_e/\omega_e = 0.1$.

## 3.2 Mode conversation at quasi-parallel propagation

As the next step, wave propagation oblique to the background magnetic field is considered. The full set of parameters used are the following: $\theta = 7^0$, $n_{s0} = 0.05$, $V_s/V_e = 4.0$, $V_e/c = 0.05$, and $G = \Omega_e/\omega_e = 0.1$. The further procedure is the following: After integrating the ordinary differential equations with the wave number $kc/\omega_e$ as parameter, the maximum amplitudes of the electromagnetic field components are determined. Results are shown in Figure 3. Whereas the longitudinal $|E_x(k)|$ remains nearly constant, maxima of the transverse electric and magnetic fields arise which are just located at the wave number where the mode crossing takes place, see Figure 1 b). For $G = 0.1$, the wave number $kc/\omega_e = 0.3$. As further seen, the maxima of the transverse magnetic and electric field components are related according to $||B\bot(k)|\sim 0.3|E\bot(k)|$.

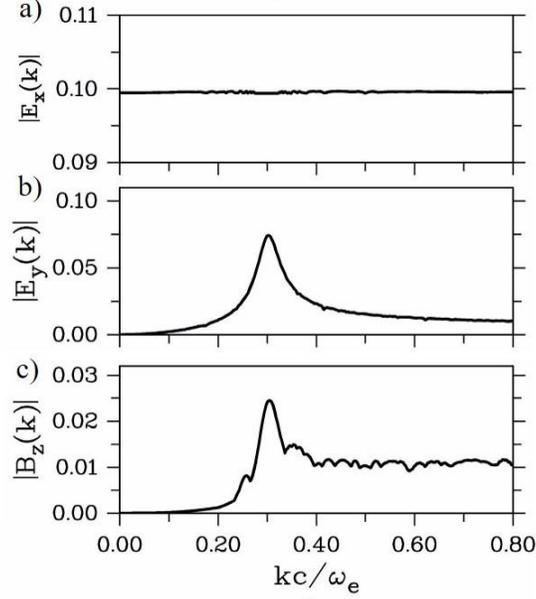

**Figure 3.** Maximum of the electric and magnetic fields components $E_x$, $E_y$ and $B_z$ versus the normalized wave number $\varkappa = kc/\omega_e$ solving the equations (1)-(11). Besides the propagation angle $\theta = 7^0$, the other parameters are the same as in Figure 2: $n_{s0} = 0.05$, $V_{s0} = 0.1$ ($V_s = 4V_e/c$, $V_e = 0.05/c$), and $G=0.1$. The electric and magnetic fields are normalized by $E_0 = cB_0$ and $B_0$, respectively. As seen, optimum coupling between the current-driven Langmuir oscillations and the electromagnetic L wave occurs at $kc/\omega_e = 0.3$, as predicted by Figure 1b.

Important information about the mode conversion is gained from the temporal evolution of the electron velocities and waves involved if the wave number of optimum coupling, $kc/\omega_e = 0.3$, is taken. As seen in Figure 4, the core and strahl velocities oscillate with the same amplitudes as for parallel propagation in Figure 2. But, amplitude oscillations appear which reflect the onset of coupling between the Langmuir wave and the adjacent left-polarized electromagnetic (L) wave. Their period is $\omega_e T \sim 700$.

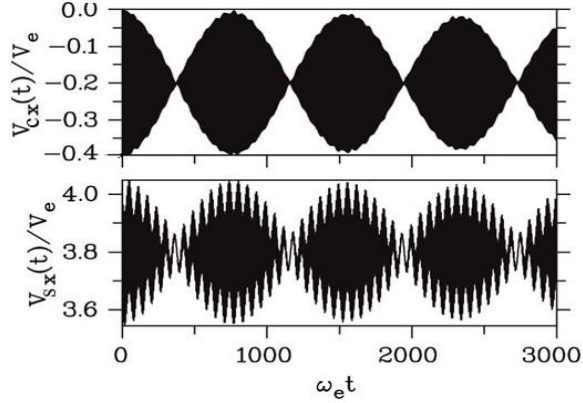

**Figure 4.** Temporal evolution of a) the core and b) the strahl velocity at the optimum wave number $kc/\omega_e = 0.3$. The same plasma parameters as in Figure 2 are used, but here for $\theta = 7^0$ (oblique propagation). The amplitude oscillations occur due to the onset of coupling between the Langmuir wave and the electromagnetic L wave, see also Figure 1.

The corresponding temporal evolution of the electromagnetic field components $E_x$, $B_y$ and $B_z$ is shown in Figure 5. All three quantities vary with the period of $\omega_e T \sim 800$ and there is a phase shift between the longitudinal electric field $E_x$ and the magnetic field components of $180^0$. Further, the (normalized) amplitude of the electric and magnetic field components are related by $|B|/|E| \sim kc/\omega_e = 0.3$, the wave number of optimum coupling for $G = 0.1$. Remarkable are the corresponding frequency spectra which are shown in the middle and right panels of Figure 5. Particularly noteworthy are the peaks at $\omega \sim 0.1\Omega_e$ in the magnetic components $B_y$ and $B_z$ of Figures 5b and 5c, respectively, which express the excitation of whistler waves.

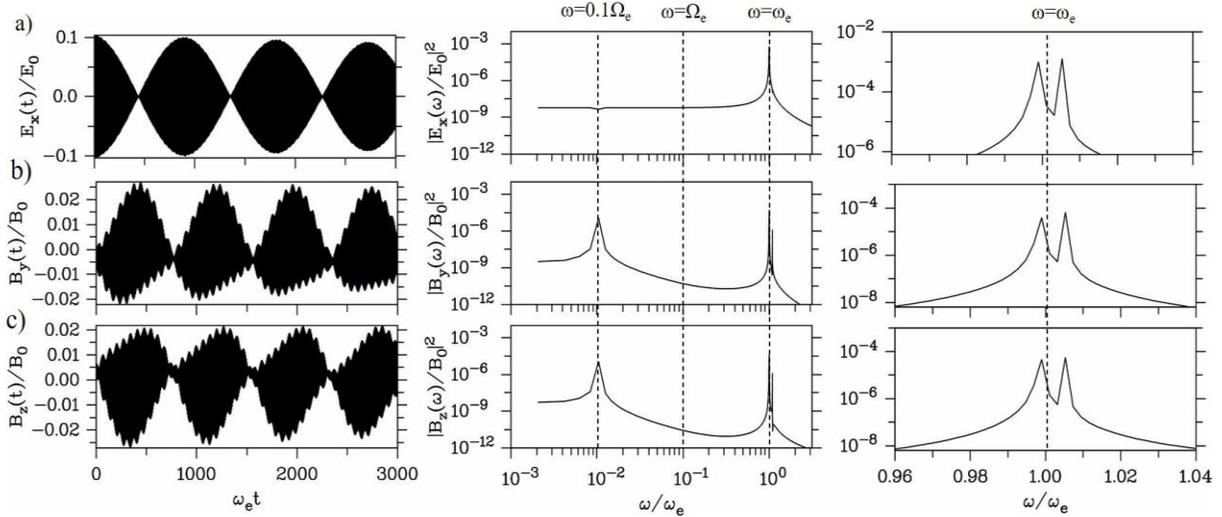

**Figure 5.** Temporal evolution and the associated frequency spectra of the electric and magnetic field components $E_x$, $B_y$ and $B_z$. The left panels show a) $E_x(t)$, b) $B_y(t)$ and c) $B_z(t)$. The middle panels represent the associated power spectra in a logarithmic scale in which the frequencies $0.1\Omega_e$, the cyclotron frequency $\Omega_e$ and the electron plasma frequency $\omega_e$ are marked by the dashed lines. In the right panels the double-peak spectra around $\omega_e$ are shown. The parameters are the following: $\theta=7^0$, $n_{s0} = 0.05$, $V_s/V_e = 4.0$, $V_e/c = 0.05$, $G = \Omega_e/\omega_e = 0.1$. The wave number of optimum mode coupling is $kc/\omega_e = 0.3$.

How the period of the amplitude oscillations occurring in Figures 5a-c changes with the propagation angle θ can be calculated from the dispersion relation of the waves involved in the mode coupling. These are the Langmuir wave and the electromagnetic L wave, see Figure 1. If the associated frequencies $\omega_1$ and $\omega_2$ are determined for a fixed ratio $\Omega_e/\omega_e$ with the wave number of optimum coupling, the period T is obtained from the difference between the two frequencies according to the relationship $T=2\pi/|\omega_1-\omega_2|$. Figure 6 shows the dependence of the period $\omega_e T$ versus θ for three values of $G=\Omega_e/\omega_e$. For the parameters of Figure 5 (G=0.1, θ=7⁰), following the dashed curve, one gets $\omega_e T\sim 800$ in full agreement with the numerical results. For nominal PSP conditions of G~0.02 one would read from Figure 6 for small propagation angles (θ≤5⁰) periods between $\omega_e T\sim 10^3$ and $\omega_e T\sim 10^4$. How this relates to the observations is discussed in the conclusions at the end.

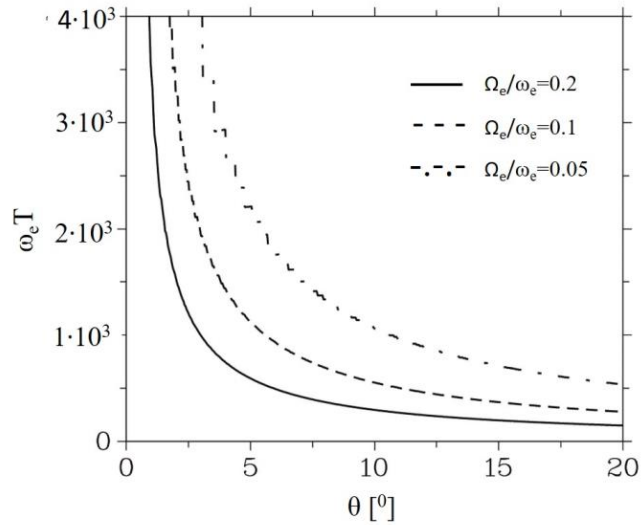

**Figure 6.** Period of the electromagnetic field oscillations ($\omega_e T$) arising from the mode coupling of the current-driven Langmuir wave and the L-wave versus the propagation angle θ for three values of $\Omega_e/\omega_e$.

Further numerical analysis has been carried out for θ≥20⁰. It shows that the optimum wave number for the generation of current-driven electromagnetic waves shifts to $kc/\omega_e\sim 1$ with the consequence that the frequency of the correlated whistler wave results from the Gendrin relation $\omega=0.5\Omega_e\cos\theta$ (Gendrin, 1961). That corresponds just to the point where phase and group velocities are the same and stationary waves (whistler oscillitons) may arise, see Sauer et al. (2002). More details follow subsequently.

In addition to the explanations at the beginning of Chapter 3, which justified the use of the uncompensated strahl at t=0, we would like to conclude this part by comparing the above described results of the fluid model for quasi-parallel propagation with those from kinetic PIC simulations. For this purpose, the power spectrum of the three quantities $E_x$, $E_z$, and $B_y$ which is presented in Figure 5 for the wave number of optimum coupling, was calculated as a function of wavenumber and frequency. The results are presented in Figure A1 of the appendix for $0\leq kc/\omega_e\leq 1$ and $0.8\leq\omega/\omega_e\leq 1.2$. The very good agreement between the results of both models clearly demonstrates that the assumptions made in the fluid model allow for an adequate description of the real conditions. This applies, on the one hand, to working with fluid equations themselves and thus neglecting kinetic effects such as wave-particle interaction, and, on the other hand, to the linearization applied, which is clearly sufficient for describing the crucial processes such as the mode coupling. The advantage of the fluid concept is the ease of handling of the equations, which are solved using simple Runge-Cutta routines.

### 3.3 Dispersion, waveforms and spectra at oblique propagation ($\theta \geq 20^0$)

In this section, the influence of the propagation angle on the spectrum of current-driven electromagnetic waves is shown. The mechanism is the same as previously described for quasi-parallel propagation. However, due to the changed dispersion behaviour, which is reflected in the different polarization of the wave modes involved, there are noticeable modifications compared to the case of quasi-parallel propagation, especially in the frequency spectrum of the excited whistler waves.

For larger propagation angles, the mode splitting is not as obvious as in Figure 1 for quasi-parallel propagation. The dispersion of the wave modes in the vicinity of the electron plasma frequency $\omega_e$ and the whistler wave ($\omega < \Omega_e$) for $\theta = 65^0$ is shown in the panels a) and b) of Figure 7. Coming from large wave numbers, the Langmuir wave (solid line) at $kc/\omega_e > 1$ goes with $k \to 0$ smoothly over into the L-wave when approaching the cutoff at $\omega \sim \omega_e - \Omega_e/2$. This altered dispersion behaviour has effects on the spectrum of waves driven by the initial electron current. In Figures 7c, d and e the maximum electric and magnetic field amplitudes ($E_x$, $E_y$ and $B_z$) versus the wave number $kc/\omega_e$ after integrating the basic equations (1)-(11) are shown as solid curves. The difference to Figure 3 for quasi-parallel propagation is clearly seen. While there both the transverse electric and magnetic fields maximize (for $G = 0.1$) at $kc/\omega_e \sim 0.3$, the maximum of $B_z$ in the present case is located at $kc/\omega_e \sim 1$. For the whistler waves, this leads to the frequency dependence $\omega \sim 0.5\Omega_e \cos\theta$ which corresponds to the Gendrin relation of whistler whose phase and group velocity coincides (Gendrin, 1961). In order to demonstrate the transition from the quasi-parallel case ($\theta \leq 15^0$) with electric and magnetic field maxima at $kc/\omega_e \sim G^{1/2}$ to the so-called Gendrin case with $kc/\omega_e \sim 1$, the corresponding dependence for $\theta = 15^0$ is shown as dashed curves. Because of the two maxima of $B_z$, two kinds of whistler waves can be expected in the transition region, namely waves with ($\omega \sim G$, $kc/\omega_e \sim 0.3$) and ($\omega \sim 0.5\Omega_e \cos\theta$, $kc/\omega_e \sim 1$). The extent to which these results are relevant for the interpretation of space measurements will be discussed later.

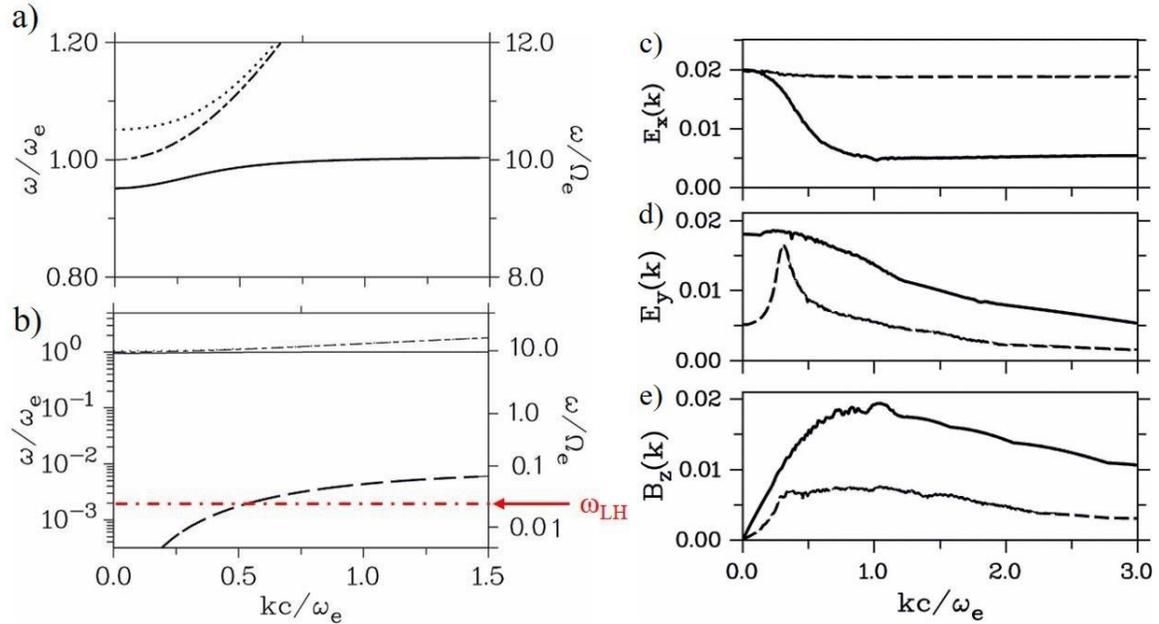

**Figure 7.** Left - Dispersion branches for oblique propagation, $\theta=65^0$: a) close to $\omega_e$ and b) below $\Omega_e$ (whistler wave) in logarithmic scale, both for $\Omega_e/\omega_e=0.1$. Right - Maximum electric and magnetic field amplitudes versus $kc/\omega_e$: c) $E_x(k)$, b) $E_y(k)$ and c) $B_z(k)$ for two propagation angles; solid line - $\theta=65^0$, dashed line - $\theta=15^0$. The strahl parameters are: $n_{s0} = 0.05$, $V_{s0} = 0.1$ ($V_s = 4V_e$ $V_e = 0.05/c$). The electric and magnetic fields are normalized by $E_0 = cB_0$ and $B_0$, respectively. Compare with Figure 3 for $\theta=7^0$ and note the occurrence of the magnetic field maximum at $kc/\omega_e$ ~ 1 for $\theta \geq 15^0$.

Choosing the optimum wave number $kc/\omega_e =1$, Figure 8 shows the velocity oscillations of the core and strahl after integrating the system of equations (1)-(11). Remarkable feature seen in panel b) is the large-amplitude oscillation in $V_{sx}$ at just the electron cyclotron frequency $\Omega_e$ which has no counterpart in the spectrum of the electromagnetic fields shown in Figure 9. In the framework of a fluid approach, this is not a surprising fact since no corresponding wave mode exists for that frequency. It would be different using a kinetic theory in which electron cyclotron harmonic modes arise. The measurements of Malaspina et al. (2020, 2021) appear to provide an experimental indication of this.

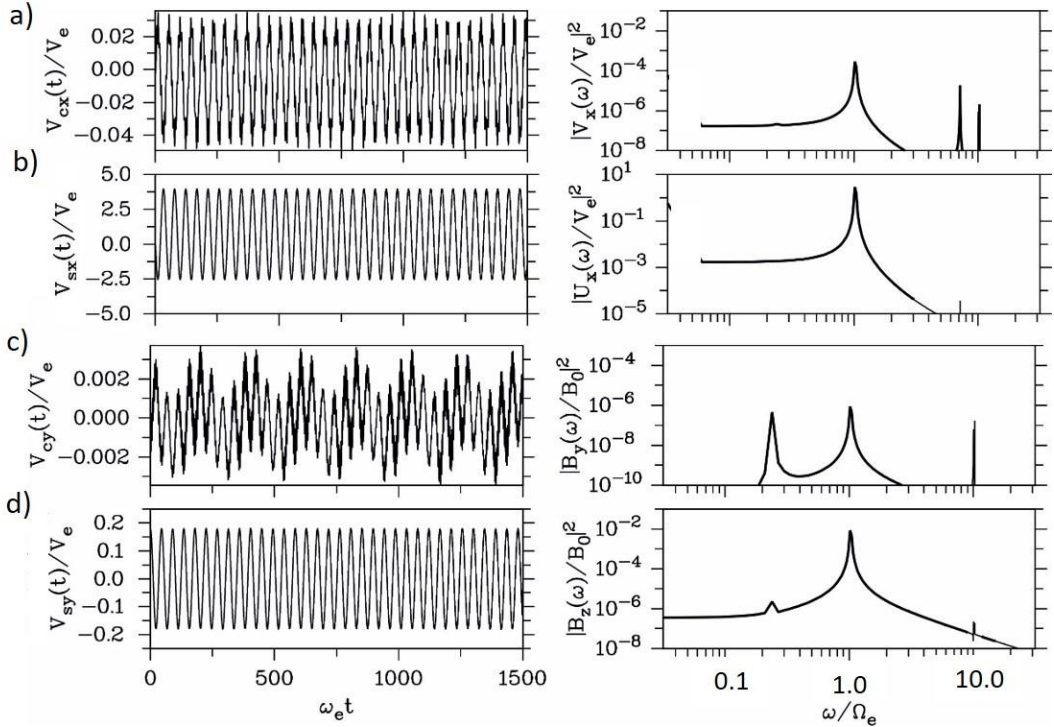

**Figure 8.** Velocity oscillations of the core and strahl electrons (left) and the associated spectra (right) for oblique propagation ($\theta = 65^0$) at the optimum wave number $kc/\omega_e = 1$. From top to bottom: $V_{cx}$, $V_{sx}$, $V_{cy}$ and $V_{sy}$.

The waveforms of the electric and magnetic field components and their associated spectra are shown in Figure 9. Two pronounced peaks are visible, one close (but not exact) to the electron plasma frequency $\omega_e$ (= $10\Omega_e$) and the other at about $0.3\Omega_e$. The latter one belongs to the whistler wave whose frequency follows roughly the Gendrin relation $\omega = 0.5\Omega_e \cos\theta$. As noted already, the peak in the spectrum of strahl velocity oscillations (Figure 8) at the electron cyclotron frequency $\Omega_e$ does not appear in that of the electromagnetic waveforms. It can be expected, however, in kinetic plasma models in which additional wave modes as magneto-acoustic and Bernstein waves are contained. With increasing propagation angle, the frequency of the whistler wave shifts more and more to the lower hybrid frequency whereby the high-frequency peak close to the electron plasma frequency remains. It should further be noted that the driven whistler wave propagates in the direction of the strahl electron flow. In the opposite direction, there are no whistler waves of the same frequency. That means, the strahl-driven whistler waves show an asymmetry with respect to the strahl flow direction. We have checked that this is different if an initially shifted halo distribution is considered. In that case, whistler waves are generated in both directions.

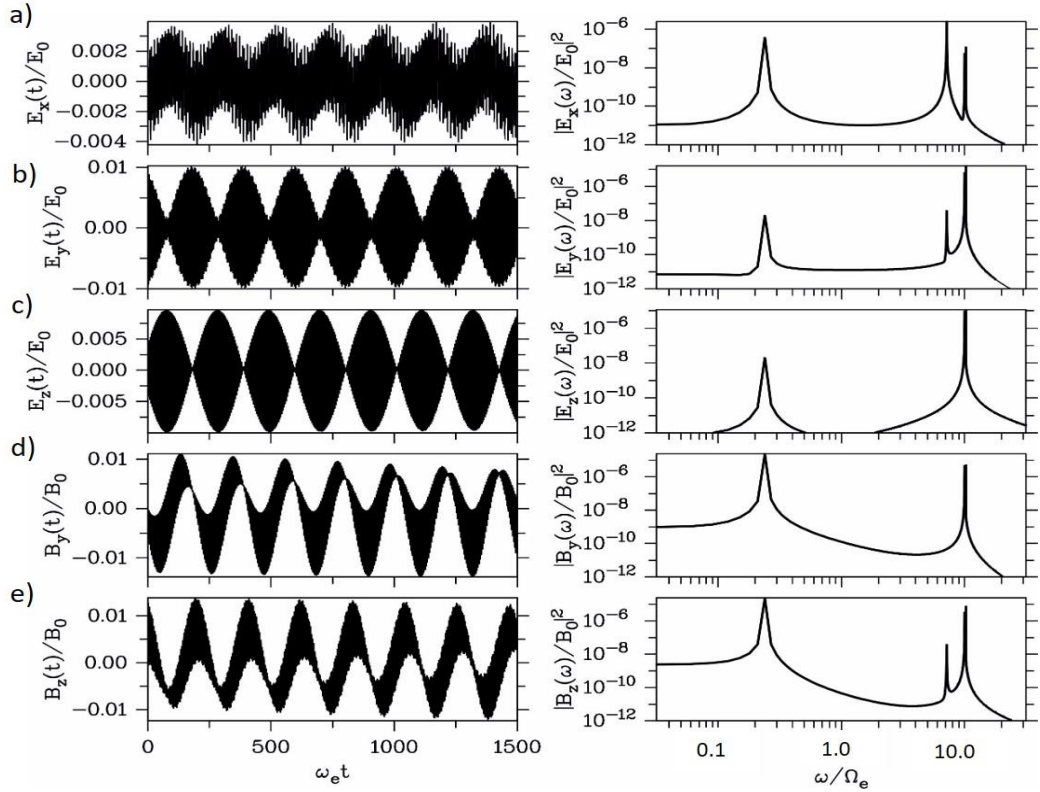

**Figure 9.** Temporal evolution (left) and the associated frequency spectra (right) of the electric and magnetic field components $E_x$, $E_y$ $E_z$, $B_y$, and $B_z$ (from top to bottom) for oblique propagation ($\theta = 65^0$) at the optimum wave number $kc/\omega_e = 1$, other parameters as in Figure 5.

## 4. Electromagnetic wave generation by a given oscillating current

In the previous section, the generation of electromagnetic radiation was considered for the case where current-driven Langmuir waves in the range $kc/\omega_e<1$ arise due to the abrupt switching on of the strahl current. As already mentioned in the beginning, this procedure was chosen to allow a comparison of the results with those of companion PIC simulations, where the same situation can be realized without extensive changes to the numerical code assuming that at $t=0$ the strahl current is not compensated by a corresponding displacement of the core.

### 4.1 Generation of type III radiation
In the following, excitation conditions will be considered that are more closely aligned with the realistic conditions of generation of type III radiation by starting with an oscillating current in a narrow frequency range which is thought to be generated by beam-plasma interaction. The general situation with respect to the wave excitation we have in mind is illustrated in Figure 10. It shows the dispersion of the existing wave modes in the frequency range around $\omega_e$. These are the two electromagnetic R and L waves and the Langmuir wave (LM). The related parameters are $\omega_e/\Omega_e=0.04$ and $V_e/c=0.02$. It is important to note the frequently discussed mode splitting in the case of oblique propagation; see also Figure 1. The dashed-dot line indicates the fictive beam mode $\omega=kV_b$ with $V_b=15V_e$. If one assumes that as a result of the instability an oscillating current with a (trigger) frequency slightly above the electron plasma frequency ($\omega=1.05\omega_e$) is generated, then there are two wavenumber ranges in which waves can be excited. On the one hand, this is the 'region of beam-instability' around $kc/\omega_e \sim 3.3$ in which the (unstable) Langmuir wave would occur. The other region of wave excitation is around $kc/\omega_e \sim 0.25$, where the L wave propagates, the polarization of which is not purely electromagnetic there due to the mode splitting. The processes directly related to the beam instability, which have been extensively investigated in the past in connection with the mechanism of plasma emission according to Ginzburg and Zhelezniakov (1958), will not be considered further here.

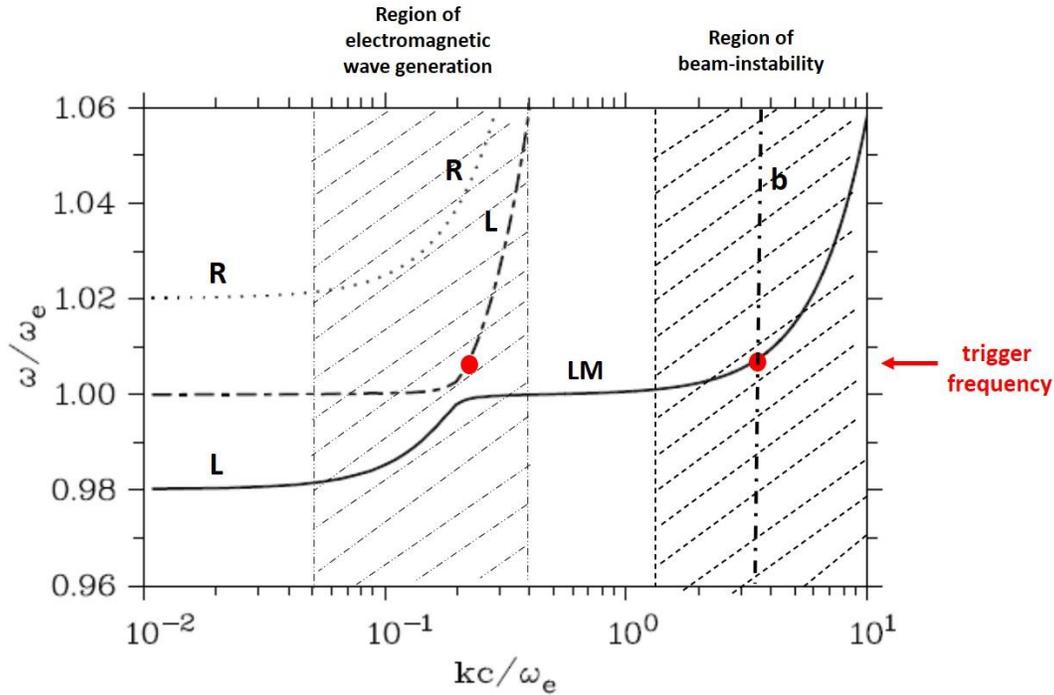

**Figure 10.** Dispersion of plasma waves for slightly oblique propagation ($\theta=10^0$) in the frequency range around the electron plasma frequency $\omega_e$. These are the electromagnetic left-hand (L) and right-hand polarized (R) wave modes and the electrostatic Langmuir (LM) mode. Further parameters are: thermal velocity $V_e/c=0.02$ and $G=\Omega_e/\omega_e=0.04$. The dashed-dot line marks the fictive beam mode $\omega=kV_b$ with $V_b=15V_e$ which may produce the oscillating current of the (trigger) frequency $\omega=1.05\omega_e$.

For our aim to describe the basic mechanism of the generation of electromagnetic waves by an electron current, the system of equations (1) – (11) can be further reduced. To do so, the equations (4), (5) and (6) for the strahl velocities are deleted and the strahl current on the right-hand side of equations (7), (8) and (9) for the electric field is replaced by a current with a pregiven temporal profile. That is: $j_s=n_s V_s \rightarrow CT$ where CT denotes the driving current. Thus, eight coupled differential equations remain from the system (1) – (11), three for the electron velocities ($V_{ex}$, $V_{ey}$, $V_{ez}$), three for the electric fields ($E_x$, $E_y$, $E_z$) and two for the magnetic field components ($B_y$, $B_z$). As for the case before, density disturbances have almost no influence on the electric fields and are therefore neglected. To mimic the onset and decay of the beam instability a current of given frequency $\omega \sim \omega_e$ is used whose amplitude increases exponentially and decays similarly after reaching its maximum.

Results of the numerical integration of the eight differential equations are shown in Figure 11 where the profile of the driving oscillating current CT(t), normalized to $CT_0=en_0c$, is plotted in panel a) with its related frequency spectrum in panel e). The assumed current has a frequency of $\omega=1.005\omega_e$, a maximum amplitude of $CT/CT_0=10^{-4}$ and growths and decays with a characteristic time of $\omega_e T \sim 100$. The other parameters are the same as before: $\theta=7^0$, $G=\Omega_e/\omega_e=0.1$. Very similar as in Figure 5, Langmuir waves are driven at $kc/\omega_e \sim 0.3$ where the mode coupling to the electromagnetic wave takes place. It results in amplitude oscillations of all electric and magnetic field components and leads to the double peak in the corresponding frequency spectra. $E_y$ and $B_z$ vary very similar as $E_z$ and $B_y$, respectively.

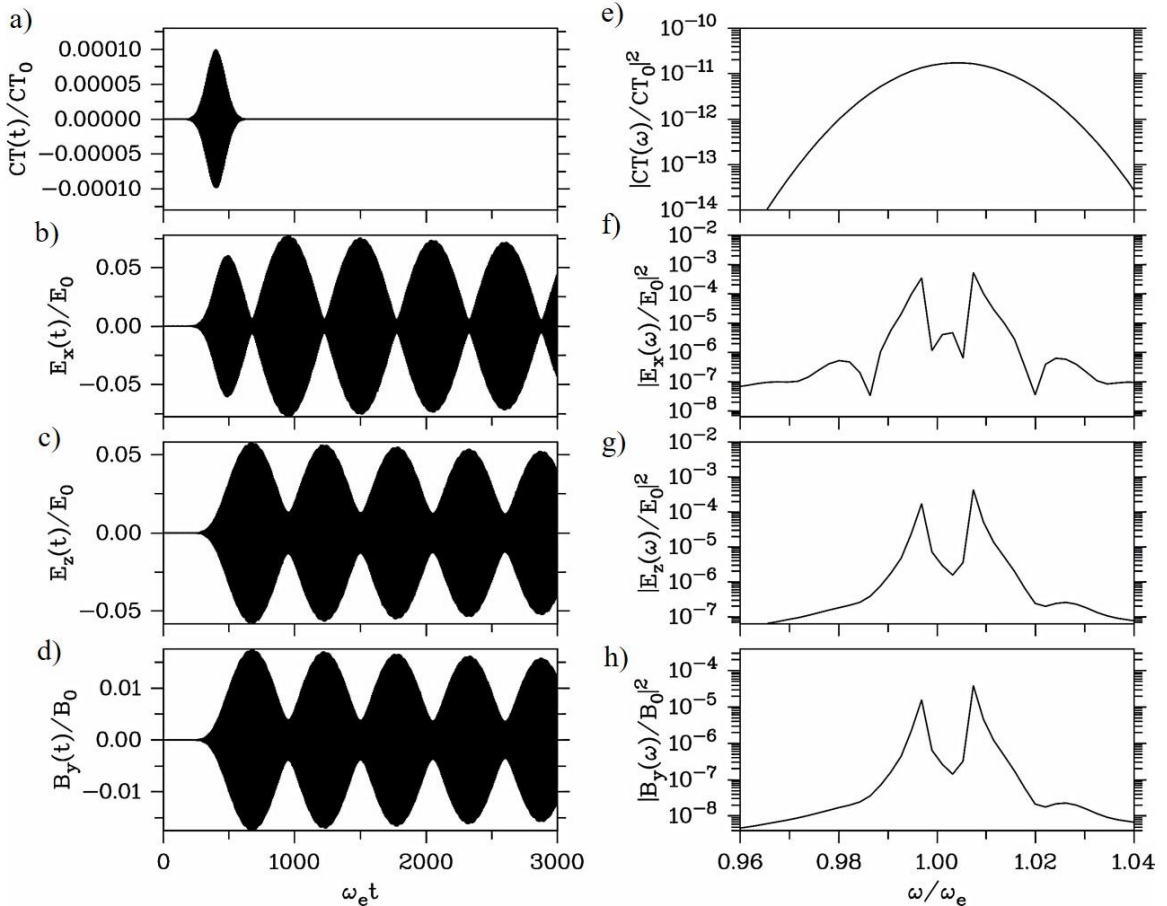

**Figure 11.** Temporal profiles of the driving current and the resulting electric and magnetic fields (left) together with the corresponding frequency spectra (right). The frequency of the driving current CT(t) in panel a) is $\omega \sim 1.005\omega_e$. The other parameters are the propagation angle $\theta=7^0$ and $G=\Omega_e/\omega_e=0.1$.

The method demonstrated here for generating electromagnetic radiation by applying an oscillating current, which mimics the processes involved in beam-plasma interaction, can be easily adapted to a wide variety of situations by adjusting the time profiles. The key factor here is the robust mechanism of excitation of Langmuir oscillations/waves by the time-varying current. The electromagnetic radiation is then generated via linear mode coupling. For example, this approach can be used to investigate how the frequency of the generating current affects the intensity of the radiation, which in turn allows conclusions to be drawn about the influence of the parameters of the underlying beam instability. Finally, it should be mentioned that the fluid approach can be easily extended to the generation of the second harmonic by calculating the nonlinear second-order currents. This is a topic of a separate study.

### 4.1 Generation of whistler waves

According to the studies in Section 3, where wave excitation by a (given) net current at time t=0 was considered, one can expect that whistler waves can also be excited by suitable current pulses. One should note here that the wavenumber of optimal excitation is determined by the current profile. As subsequently will be seen, sharper pulses will lead to larger wave numbers and thus to higher frequencies. The measure is the half-width ($T_o$) of an assumed Gaussian profile. In Figure 12 a) the selected current pulse with $\Omega_e T_0=20$ and a maximum amplitude of $CT/CT_0=0.025$ is shown. The corresponding frequency spectrum of the pulse in panel e) indicates that whistler waves up to about $0.2\Omega_e$ can be excited. The numerical integration of our system of five differential equations yields, as shown in Figure 12, the temporal evolution of the electromagnetic fields $E_x$, $E_y$ and $B_z$ (left) and the resulting frequency spectra (right). $kc/\omega_e=0.3$ has been determined as the optimum wave number. As can be clearly seen from the temporal profiles of the $E_z$ and $B_y$ fields in panels c) and d), the whistler wave is very coherent, which is also clearly reflected in the pronounced peak at $0.1\Omega_e$ in the spectra of panels e) and f). Due to the linear relationship between the strength of the current pulse and the intensity of the generated electromagnetic field, these can be easily scaled. How these results relate to previous and current space craft observations (e.g. Agapitov et al., 2020) is discussed at the end.

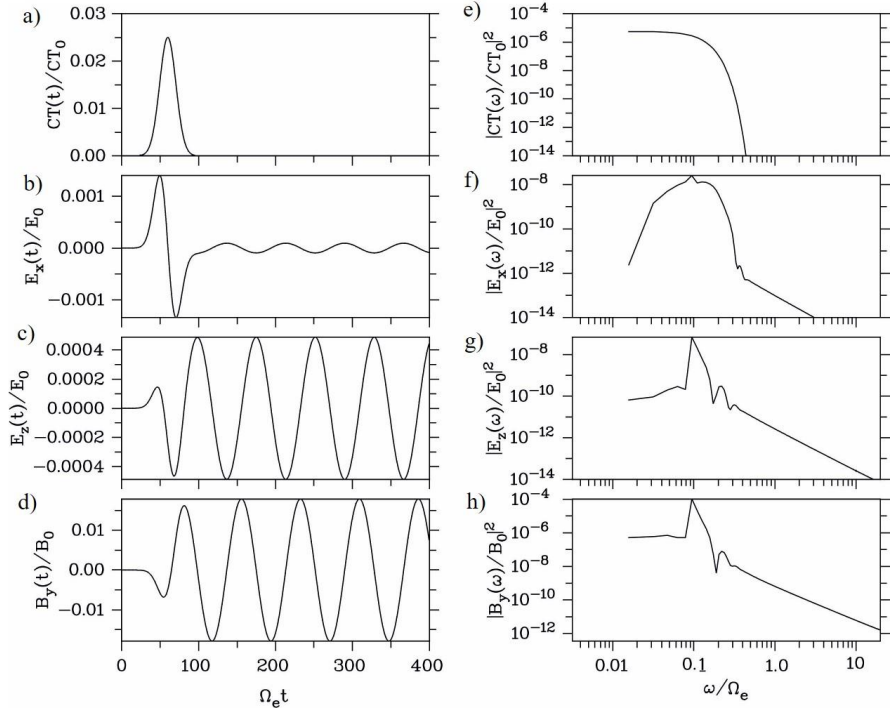

**Figure 12.** Excitation of whistler waves by a current pulse. Temporal profiles of the driving current and the resulting electric and magnetic fields (left) together with the corresponding frequency spectra (right). The characteristic growth and decay time of the driving current CT(t) in panel a) is $\Omega_e T_0 \sim 20$. The other parameters are the propagation angle $\theta=10^0$, $G=\Omega_e/\omega_e=0.1$ and the (optimum) wave number $kc/\omega_e \sim 0.3$. As seen in the panels g) and h), a whistler wave with the frequency $\omega \sim 0.01\omega_e=0.1\Omega_e$ is driven by the temporal variation of the driving current.

How the (optimum) wave number and the associated frequency varies with the temporal profile of the triggering current is presented in Figure 13. Panel a) shows three Gaussian profiles with different half-widths of 1: $\Omega_e T_0=4$, 2: $\Omega_e T_0=15$ and 3: $\Omega_e T_0=40$. Panel b) shows the (maximum) magnetic component $B_y$ versus the wave number $kc/\omega_e$ after integrating the governing differential equations. The dotted curve represents the dispersion relation of the triggered whistler wave. Starting with the sharp profile with $\Omega_e T_0=4$ (curve 1), the associated optimum wave number is at $kc/\omega_e \sim 0.75$ which corresponds to the excitation of a whistler wave with the frequency $\omega \sim 0.35\Omega_e$. Smoother current profiles shift the optimum wave numbers to smaller values. For the Gaussian profile with $\Omega_e T_0=40$ (curve 3), the maximum of $B_y(k)$ is located at $kc/\omega_e \sim 0.2$ with the frequency $\omega \sim 0.04\Omega_e$. The maximum (optimum) wave number is $kc/\omega_e=1$. Thus, current-driven whistler waves may occur up to a maximum frequency of half the electron cyclotron frequency $\Omega_e$, i.e. $\omega \leq \Omega_e/2$.

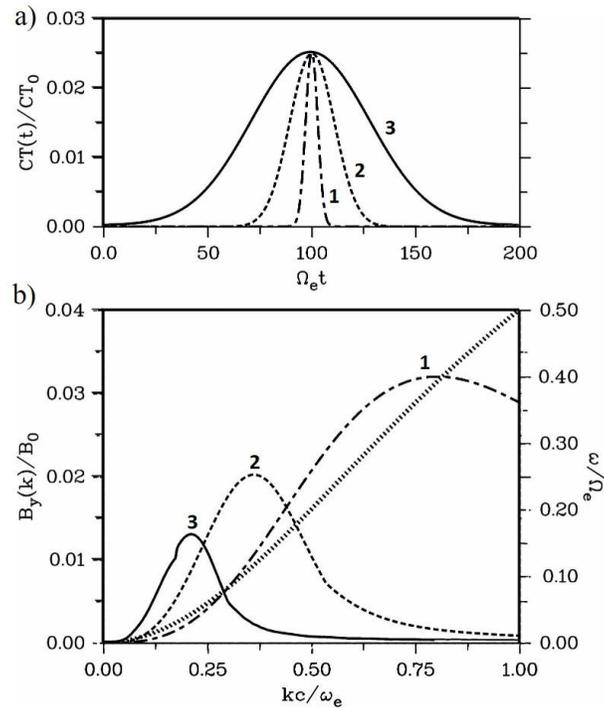

**Figure 13.** Whistler waves triggered by Gaussian current profiles. a) Gaussian current profiles $CT(t)$ with different half-widths $T_0$. 1: $\Omega_e T_0=4$, 2: $\Omega_e T_0=15$ and 3: $\Omega_e T_0=40$. The current is normalized to $CT_0=en_0c$. b) Magnetic field component $B_y/B_0$ versus the normalized wave number $kc/\omega_e$. The selected propagation angle is $\theta=10^0$. Smoother profiles shift the optimum wave number to smaller values that means to smaller whistler frequencies.

### 4.2 Joint phase-stable excitation of whistler and Langmuir waves

Inspired by previous and current publications discussing the simultaneous occurrence of Langmuir and whistler waves and their possible origin (e.g. Kennel et al, 1980; Jagarlamudi et al. 2021), two examples are considered. In the first case, a current pulse is assumed that is sufficiently short to ensure excitation of both waves, similar to a non-compensated net current at t=0 in Figure 5 of Section 3.1. Subsequently, a broader current pulse is considered, the same as in Figure 12 a), but additionally superimposed by stochastic fluctuations.

In Figure 13, results are shown for the case of a short pulse with a characteristic width of $\omega_e T_0=5$. Panels b), c), and d) show the temporal evolution of the electromagnetic components $E_x$, $E_z$, and $B_y$ in the interval $0<\omega_e t<100$. The entire evolution up to $\omega_e t=2000$ is plotted in the corresponding panels in the middle. In the right-hand panels, the corresponding spectra clearly indicate that whistler waves and Langmuir waves are excited simultaneously, with the longitudinal field ($E_x$) of the Langmuir wave being the strongest in the considered case, selecting $kc/\omega_e=0.3$. Larger wavenumbers lead to a moderate increase in the whistler wave amplitude.

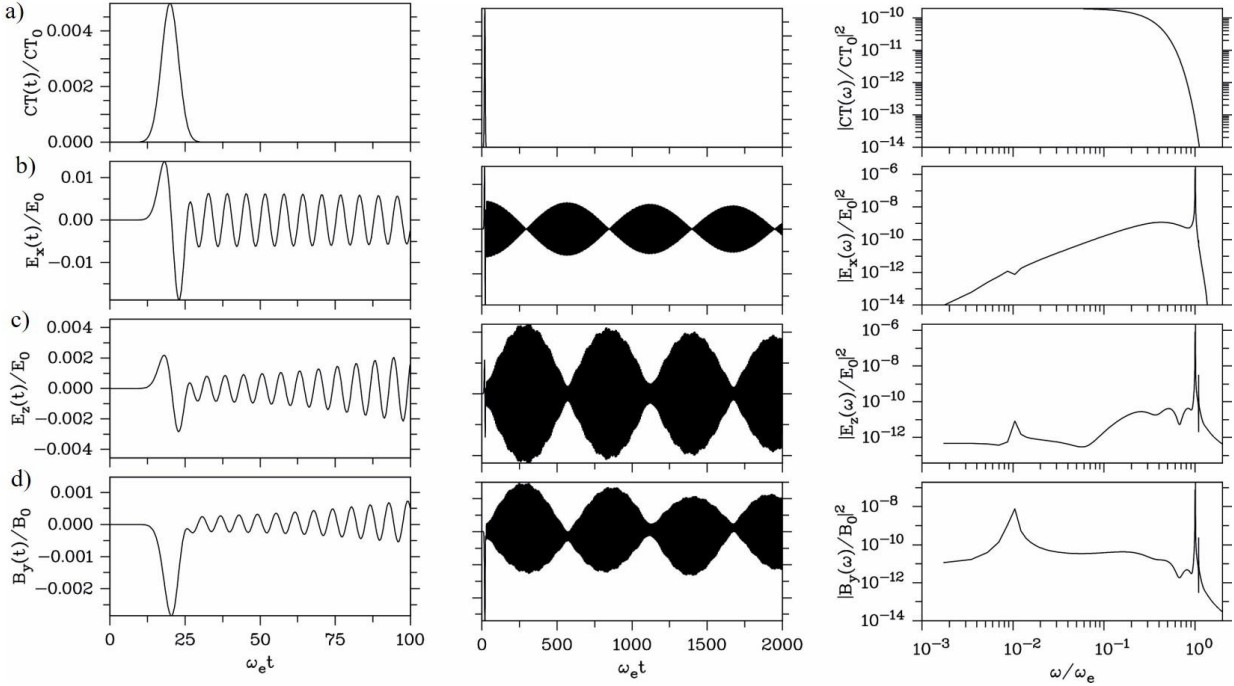

**Figure 13.** Simultaneous excitation of Langmuir and whistler wave by a short current pulse with a half-width of $\omega_e T_0=5$. The left panels represent the temporal variation of a) the current pulse, b) the longitudinal electric field $E_x$, c) the transverse electric field $E_z$ and d) the magnetic field component $B_z$ in the time interval $0\leq\omega_e t\leq100$. The same quantities over up to $\omega_e t=2000$ are shown in the middle panels. The right panels show the related power spectra. The selected wave number is $kc/\omega_e=0.3$.

Next, the consequences resulting from the superposition of a current pulse with stochastic fluctuations are investigated. For this purpose, the same current pulse as in Figure 11a) is considered, superimposed with stochastic fluctuations of a given amplitude of $\delta j/j_0=5\cdot 10^{-4}$. As the corresponding power spectrum in panel 14e) shows, these current fluctuations lead to contributions in the high-frequency range around the electron plasma frequency $\omega_e$. This is associated with the excitation of Langmuir waves, which ultimately generate type III electromagnetic fields with the described amplitude oscillations as a result of mode coupling. The high-frequency oscillations are also visible in the magnetic component $B_y$ in panel d). How the superposition appears in detail depends on several parameters, whereby the ratio $G=\Omega_e/\omega_e$ determines the frequency of the whistler wave and the propagation angle $\theta$ essentially influences the amplitude oscillations in the high-frequency range. Thus, the phase position of the minima of $E_z$ with respect to the magnetic component $B_z$ is not determined solely by the whistler wave as suggested e.g. by Li et al. (2017), but varies with the parameters G and θ according to the presented generation mechanism by which the high- and low-frequency wave are jointly current-driven.

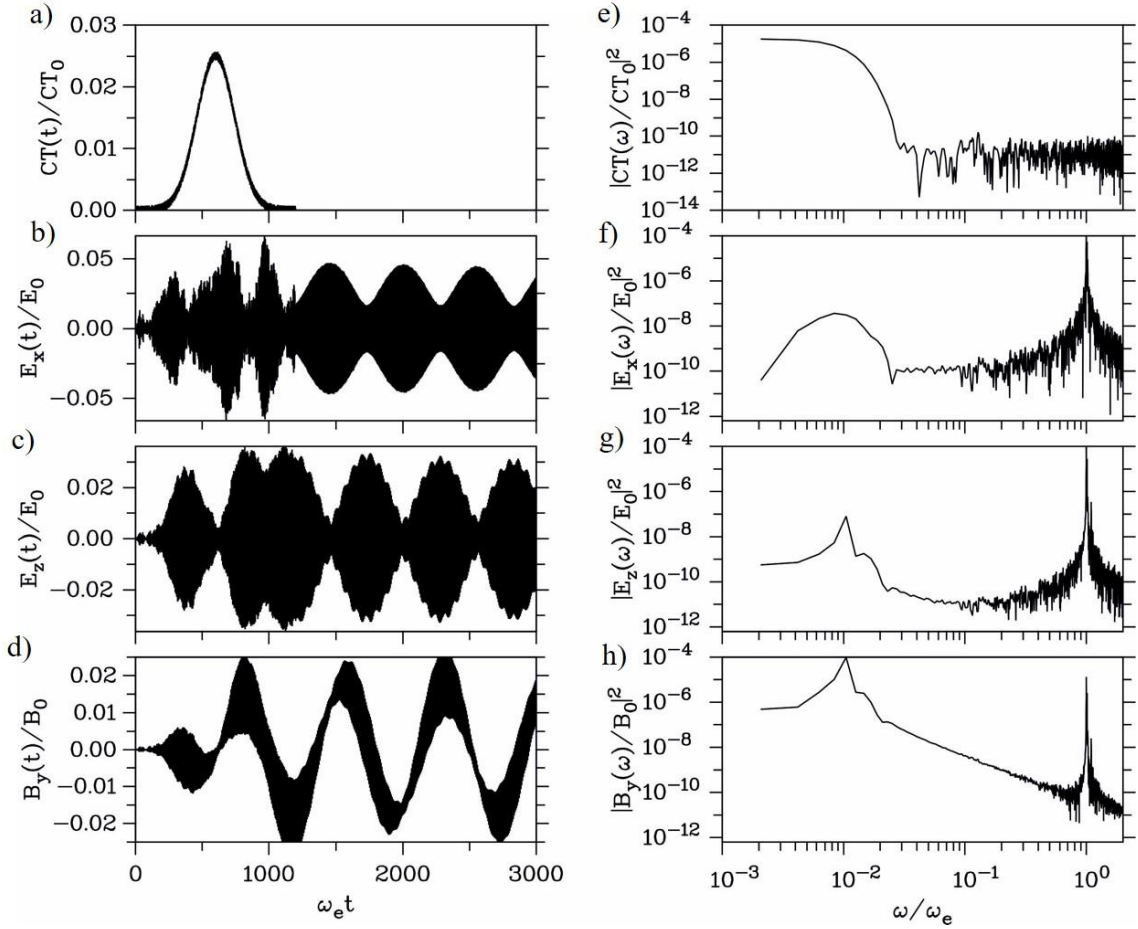

**Figure 14.** Simultaneous excitation of a whistler wave and a high-frequency wave at $\omega_e$, by the same current pulse as in Figure 11, but superimposed by stochastic fluctuations with $\delta j/j_0 \sim 10^{-3}$. As obvious, in addition to the (large-amplitude) whistler wave at $\omega \sim 0.1\Omega_e$, a type III electromagnetic wave at $\omega_e$ is generated by the arising quasi-longitudinal electric field $E_x$ by which the high-frequency electromagnetic components are generated via linear mode coupling.

## 5. Summary and Discussion

The aim of our paper was to demonstrate that the generation of type III radiation generation can be described in a simple fluid model. Solving the linearized system of fluid-Maxwell equations of a core-strahl plasma in the approximation of cold electrons, in Section 2 it has been shown that the initial current of the strahl drives Langmuir oscillations/waves at the electron plasma frequency. They manifest as velocity oscillations of both the core and strahl electrons. According to Equation (17), the amplitude of the associated electric wave field is proportional to the strahl current. At oblique (quasi-parallel) propagation, the coupling of the Langmuir mode with the electromagnetic L mode leads to amplitude oscillations of all electric and magnetic field components versus time by which a spectral double-peak at $\omega_e$ arises. The wave number of optimum coupling depends on the ratio of $G=\Omega_e/\omega_e$. It shifts with G according to $kc/\omega_e \sim [G/(1+G)]^{1/2}$. In addition to the electromagnetic waves near $\omega_e$, a whistler wave is driven at frequency given by the wave number of optimum coupling. For $G\sim0.1$, as usually taken, the resulting frequency of the whistler wave is $\omega\sim0.1\Omega_e$. With decreasing G, it moves to smaller values. The formalism used is similar to the response theory of current-driven waves briefly described in Sauer et al. (2019). At oblique propagation ($\theta \geq 25^0$), the optimum wave number shifts to $kc/\omega_e\sim1$ with the consequence that the frequency of the current-driven whistler wave corresponds to the Gendrin relation $\omega\sim(\Omega_e/2)\cos\theta$.

In Section 3, instead of considering a strahl net current, the ignition of Langmuir waves in the wave number range of electromagnetic waves (see Figure 10) by current oscillations of given frequency and their transformation into type III radiation has been investigated. In this way, it is possible to simulate radiation generation during beam-plasma interaction by choosing relevant current profiles. It has been shown that the amplitude of the triggered Langmuir waves has a maximum if the frequency of the oscillating current is close to the electron plasma frequency $\omega_e$. It decays for larger or smaller frequencies. The wave number of optimum amplitude is always given by $kc/\omega_e \sim [G/(1+G)]^{1/2}$. For the triggering of whistler waves only relatively soft changes of the driving current are required. They can be excited by currents whose characteristic growth or decay times ($T_0$) are smaller than $\omega_e T_0 \sim 2\pi/G$. A further interesting result concerns the joint excitation of Langmuir and whistler waves if the whistler wave driving current is superimposed by stochastic fluctuations.

The observations of whistler waves in the solar wind and the simultaneous occurrence of Langmuir waves and type III radiation have been a topic of debate for a long time. Early magnetic signatures of Langmuir waves have been observed by Scarf et al. (1970, 1971). Examples of correlated Langmuir and (chorus) whistler waves were also reported in Kennel et al., (1980), Reinleitner et al. (1982, 1984), Moullard et al. (1998, 2001), Stenberg et al, (2005), Vaivads et al. (2007), Li et al. (2017, 2018) and in recent PSP observations, e.g., by Jagarlamudi et al. (2021) and Rasca et al. (2022), mostly related to magnetic dips at the boundary of switchbacks. The simultaneous observation of whistler and Langmuir waves at magnetopause crossings by the four CLUSTER spacecrafts (Stenberg et al., 2005) seems to be related to the existence of electron-scale density jumps. A huge variety of space measurements concern the appearance of whistler waves alone for which the obtained parameters can nicely be compared with the prediction of our theoretical approach. The recurring observation of quasi-parallel whistler waves in the frequency range around $0.1\Omega_e$ under otherwise very different conditions is striking. Examples can be found in Dubinin et al. (2007), Agapitov et al. (2020), Tong et al. (2019), Jagarlamudi et al. (2021), Cattell et al. (2020, 2021), Dudok de Wit et al. (2022), Karbashewski et al. (2023), Froment et al. (2023), Jiang et al. (2024) and Colomban et al.(2025). A further support of our theoretical model of current-driven waves comes from observations in which both frequency and wave number have been measured. Moullard et al. (2001) found obliquely propagating ($\theta \sim 20^0$) whistler waves in WIND data with frequencies of $\omega \sim 0.35\Omega_e$ whose wave numbers are given by $kc/\omega_e \sim 0.9$. These waves obviously belong to the class of oblique whistler waves whose frequency follows the Gendrin relation. A similar type has been analysed by Cattell et al. (2020) using STEREO waveforms. The median propagation direction of the observed coherent whistler waves with frequencies of $\sim0.2\Omega_e$ was $\theta \sim 65°$, in accordance with the Gendrin relation if $kc/\omega_e \sim 1$ is assumed. In high-resolution data from the MMS mission (Jiang et al., 2024), on the other hand, for the quasi-antiparallel propagating whistler waves with a power peak at ~0.1 $\Omega_{ce}$, a wave number of $kc/\omega$ ~0.4 has been determined. Particularly noteworthy in this context is the recent paper by Nair et al. (2025), which reports on the PSP observation of very coherent whistler waves, which begin very abruptly at frequencies $\omega$~$0.04\Omega_e$ and extend to about $0.2\Omega_e$. Quasi-parallel whistler waves in a similar frequency range have recently been reported by Colomban et al. (2025). Interesting measurements involving whistler waves associated with density variations are described in the work of Moullard et al. (2002). The observed frequency bands at $\omega$~$0.05\Omega_e$ and $\omega$~$0.5\Omega_e$ during CLUSTER's passage through the plasmapause are, we believe, caused by electron currents induced by the density jumps. The task for the future remains to estimate the currents that arise during abrupt parameter changes at shocks, magnetic holes, switchbacks, and other discontinuities in the plasma triggering the observed whistler waves.

The investigated correlated excitation of Langmuir and whistler waves by an initial current is also a reason to re-evaluate analyses of instabilities. As we have seen, the occurrence of whistler waves in the frequency range around $0.1\Omega_e$ is obviously not due to an instability, but is apparently caused by a current pulse that simultaneously may trigger electromagnetic waves at the electron plasma frequency. Finally, we would like to point out that the mechanism of current-driven waves could provide an explanation for the formation of chorus waves that differs from previous considerations. In particular, the idea arises that the repetition time of the chorus wave is connected to the amplitude oscillations of the Langmuir/Z wave. In the rhythm of these oscillations, a whistler wave is generated at $\omega \sim 0.1\Omega_e$, whose frequency subsequently shifts to higher values until the steady state at $\Omega_e/2$ is reached. Experimental evidence for such an explanation is seen in the measurements of Li et al. (2017, 2018)

A phenomenon observed in many studies on type III radiation associated with beam-generated Langmuir waves concerns the amplitude oscillation of the electric field components, which leads to the appearance of double peaks in their spectra, e.g. Malaspina et al. (2011). In most interpretations, this effect is explained by the parametric decay of Langmuir waves, although some considerations of their conversion to radiation have also been incorporated, e.g., Graham and Cairns (2013) and references therein. In our fluid model, the amplitude oscillation arises directly from the generation mechanism at the point of optimal coupling, which involves two wave modes with mixed polarization. As explained in connection with Figure 6, the period of this oscillation can be easily determined as a function of $G=\Omega_e/\omega_e$ and the propagation angle $\theta$. For the solar wind conditions with G~0.03 one gets from Figure 6 a medium period of $\omega_e T$~$10^3$ which leads to T~1ms if $\omega_e$~$10^6$ s$^{-1}$ ($n_e$~150cm$^{-3}$ from PSP measurements (Mozer et al., 2024) is taken. Amplitude oscillations of this order have been reported by Larosa et al. (2022) and Dudok de Wit et al. (2022). Closer to the Earth, in electric field data from STEREO (Graham and Cairns, 2013) larger periods of T~(2-5)ms have been seen.

With regard to electric field intensities, we refer to the paper of Mozer et al. (2024) in which a strong type III burst has been analysed. Measured electric field amplitudes of about 600 mV/m are in our view related to the initial strahl current. If we take a plasma with G = 0.02, $V_e/c$ = 0.01 and $B_0$ ~ 80 nT, with the assumed parameters of the strahl $n_s/n_0$= 0.02 and $V_s/V_e$= 3, one gets according to Equation (17) an electric field of E~1 V/m. In the view of the rough estimations, this seems an acceptable agreement.

Another important aspect concerns the scattering of electrons by oblique whistler waves. From the observation of broadening the strahl at the time of whistlers, they are considered as one of the main candidates for explaining the scattering of strahl electrons into the halo at increasing radial distances from the Sun, e.g., Cattell et al. (2021) and Jagarlamudi et al. (2021). According to our theory, this picture has to be revised. The oscillations of the minor electron populations (halo and/or strahl) and the accompanying electromagnetic waves form an intrinsic coupled system which is triggered by the electron current. The onset of wave-particle interaction which is associated with electron heating leads finally to the termination of the electric field and particle oscillations. This, of course, can only be described within a kinetic theory, as presented in the companion paper by means of PIC simulations.

We would like to point out that in this work we have only considered the triggering of electromagnetic waves that occur in the wavenumber range $kc/\omega_e < 1$. The focus was on the conversion of Langmuir waves into electromagnetic radiation and the associated excitation of whistler waves. The assumption of a cold plasma was therefore justified. However, it should be noted that under more realistic conditions, the presence of a beam/strahl distribution with the velocity $V_s$ is associated with an electro-acoustic mode, which interacts with the Langmuir mode and thus leads to mode splitting at the wave number $k\lambda_D=kV_e/\omega_e\sim V_e/V_s<1$. This is indicated in Figure 10 by the region of beam instability; s.a. Malaspina et al (2011). Similar to before, coherent electrostatic waves are then triggered by the strahl current, whereby, equivalent to the whistler wave in the region of electromagnetic wave generation, an ion-acoustic wave is simultaneously driven in the low-frequency range. Investigations of this kind are described in an earlier paper (Sauer et al., 2017) in which measurements of the electric field on board the WIND spacecraft are related to the currents of plateau-like electron distributions.

With these considerations in mind, it is logical to conclude that past simulations investigating the generation of type III radiation (Kasaba et al., 2001; Rhee et al., 2009; Thurgood and Tsiklauri, 2015, Henri et al., 2019; Zhang et al., 2022; Chen et al., 2022, Lazar et al. 2025) must be reinterpreted. In our opinion, the evidence of beam-generated Langmuir waves and electromagnetic radiation observed in the simulations does not necessarily allow the conclusion that the classical plasma emission according to the model of Ginzburg and Zhelezniakov (1958) has been confirmed.The two-stage model for the conversion of beam-generated Langmuir waves with $k\lambda_D\leq 1$ into radiation implies a one-sided focus on the processes of parametric decay and wave coalescence involving ion-acoustic and backscattered Langmuir waves which both are strongly damped in plasmas with $\beta_e\sim 1$. The simultaneous participation of Langmuir waves in the range $k\lambda_e=kc/\omega_e\leq 1$ has not been considered. This is partly due to the fact that this long-wavelength range was not sufficiently resolved in the 2D simulations.

Also noteworthy are the conclusions we obtained in our model regarding radiation generation in non-magnetized plasmas, which have not been discussed here. To do this, the transition $G=\Omega_e/\omega_e\to 0$ must be introduced into the governing equations. The decrease in radiation intensity observed in PIC simulations by Thurgood and Tsiklauri (2015), Zhang et al. (2022), and Lazar et al. (2025) for the case where the frequency of the unstable Langmuir wave is smaller than $\omega_e$, can be easily reproduced. It is a consequence of the fact that the excitation of Langmuir waves only occurs to a reduced extent or even not at all when the frequency of the triggering current drops below the electron frequency.

We would like to emphasize that the theory of current-driven Langmuir oscillations presented here represents a completely new approach to the generation of electromagnetic waves in plasmas. In fully contrast to excitation by instabilities, in which only resonant particle groups are involved, in the described model the entire coupled system of particles and waves takes part in the interaction. As a result, the oscillations do not decay with the saturation of an occurring instability, but are maintained far longer because of the nearly undamped Langmuir waves in the wave number range $kc/\omega_e \leq 1$. In addition, it should be pointed out that the presented fluid approach of triggered plasma wave emission can be easily extended to other wave modes than considered here. We are therefore optimistic that the unique cascade of triggered ion-acoustic waves measured by Mozer et al. (2021, 2022, 2023) can be interpreted by a modified formalism of current-driven waves that includes the previously neglected ion dynamics. In this way, kinetic Alfvén waves are taken into account, which we believe play an essential role in the formation of the observed wave cascade of low- and high-frequency waves. A very similar approach can also be taken investigating EMIC waves. When applying the formalism of current-driven waves to multi-ion plasmas, the system of required differential equations can be significantly reduced by assuming massless electrons. Maximum emission is expected to occur at the ω-k points where mode coupling takes place and stationary waves with phase equal to group velocity, called EMIC oscillitons (Sauer and Dubinin, 2022) exist

**Appendix: Comparison of fluid approach and PIC simulations**

A comparison is made between the two approaches, the fluid model and PIC simulations, which are applied to the same problem of electromagnetic wave generation by a sudden onset of the strahl current in a core-strahl plasma. For this purpose, first a short description of the PIC code should be given: One-dimensional PIC simulations are performed to investigate the consequences of a sudden onset of electron current with respect to generation of Langmuir waves with $kc/\omega_e < 1$ and their conversion into type III radiation. In these PIC simulations, both electrons and ions are treated as simulation particles (of different masses and signs of charge). As before, the simulations are one-dimensional with the simulation domain along a direction at an angle with respect to the background magnetic field. Following the wave dispersion analysis in Section 2.1, the size of the simulation domain is set to be $L = 4000\lambda_e$, where $\lambda_e = c/\omega_e$ is the electron skin length. There are

51200 cells and 1000 simulation particles per cell to represent each particle population. The simulation time step is .

The plasma parameters have been chosen same as before except that core electrons, strahl electrons and ions have a finite temperature. The core electrons and ions have Maxwellian distributions of the same temperature. Using an electron temperature according to $V_e/c=0.05$ and the same ratio of electron cyclotron frequency to electron plasma frequency as in the fluid model of $G=\Omega_e/\omega_e=0.1$, the electron plasma beta of the core electrons is $\beta_e \sim 0.7$. As in the fluid approach, triggering of the Langmuir waves is realized by the sudden onset of the strahl current. The strahl electron density is $n_s/n_0 = 0.05$; strahl electrons initially have a drift Mawellian distribution with the same thermal velocity as core electrons and a drift velocity of $V_s=4V_e$. This configuration should roughly adapt the electron distribution function which has been measured by Mozer et al. (2024) in connexion with an intensive type III radio burst.

For the comparison of both approaches, the power spectra of the magnetic and electric fields $B_y$, $E_z$ and $B_y$, respectively, are presented (from top to bottom) versus k and ω . The plots in the left panels are from the fluid model in Section 2.3, the right ones are from the PIC simulation. As can be seen, there is excellent agreement between the two models, although they differ fundamentally in their theoretical approaches and also require very different numerical effort. While the temporal evolution of the electromagnetic fields can be determined in seconds using the linear fluid equations, the PIC simulations run for hours. The agreement in the intensity of the electromagnetic radiation makes it clear that the assumption of a cold plasma in the fluid description is a good approximation, considering that the PIC simulations used a warm plasma with $\beta_e \sim 0.7$.

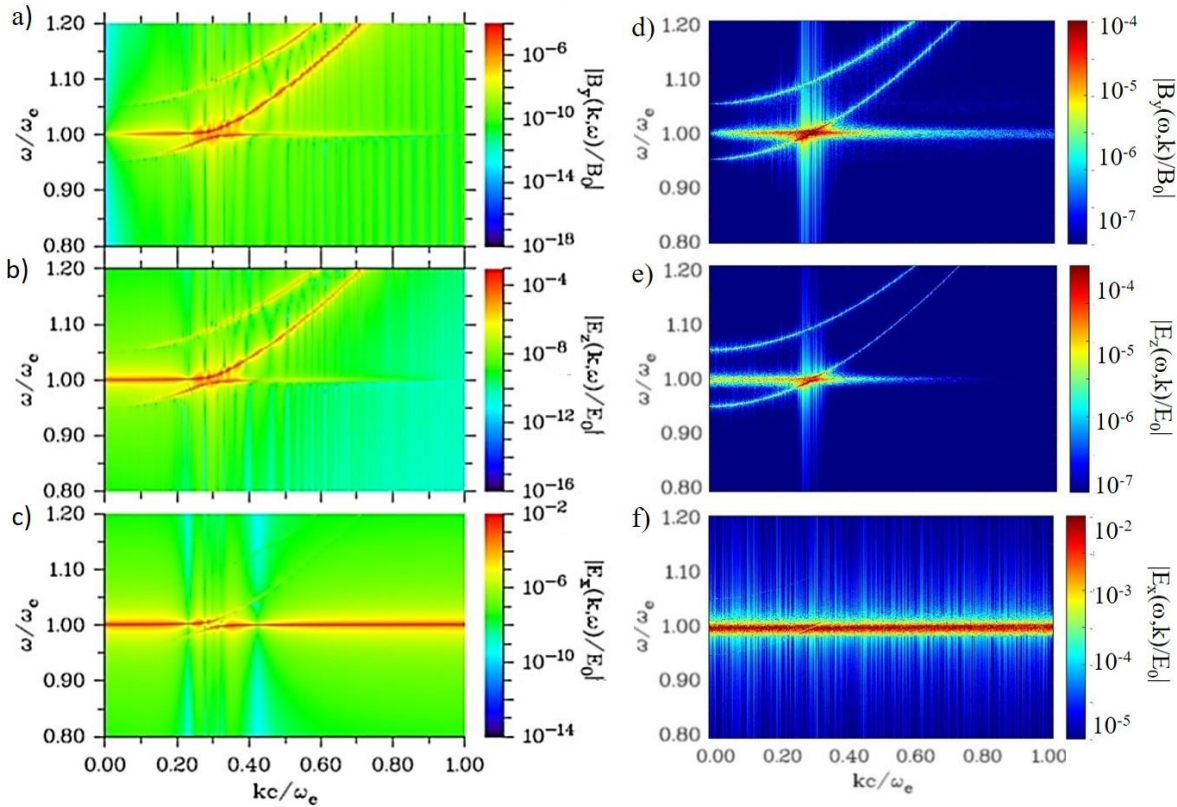

**Figure A1.** Power spectra of the magnetic $B_y$ (a, d) and the two electric fields $E_y$ (b. e) and $E_x$ (c, f) which are triggered by the net current of the strahl at t=0 with $j_s/j_0$=0.01, $j_0$=$en_0c$. In the PIC simulation the following parameters have been used: thermal electron temperature of $V_e/c$=0.05, strahl density and velocity of $n_s/n_0$=0.05 and $V_s$=$4V_e$, respectively. The results in the left panels are from the fluid approach; the right panels belong to the PIC simulations. The results are nearly identical. In both cases, the electromagnetic radiation is generated at $kc/\omega_e$~0.3 where the current-driven Langmuir waves ($\omega$~$\omega_e$) couple with the electromagnetic L wave. The dispersion of the three modes involved, the R, L and Langmuir wave, are clearly marked and correspond to that of Figure 1.